\documentclass[aps,prc,preprint,superscriptaddress,groupedaddress,
showpacs,floatfix]{revtex4}
\usepackage{graphicx}

\begin{document}

\title{Superscaling, Scaling Functions and Nucleon
Momentum Distributions in Nuclei}

\author{A.~N.~Antonov}
\affiliation{Institute for Nuclear Research and Nuclear Energy,
Bulgarian Academy of Sciences, Sofia 1784, Bulgaria}

\author{M.~K.~Gaidarov}
\affiliation{Institute for Nuclear Research and Nuclear Energy,
Bulgarian Academy of Sciences, Sofia 1784, Bulgaria}

\author{M.~V.~Ivanov}
\affiliation{Institute for Nuclear Research and Nuclear Energy,
Bulgarian Academy of Sciences, Sofia 1784, Bulgaria}

\author{D.~N.~Kadrev}
\affiliation{Institute for Nuclear Research and Nuclear Energy,
Bulgarian Academy of Sciences, Sofia 1784, Bulgaria}

\author{E.~Moya de Guerra}
\affiliation{Instituto de Estructura de la Materia, CSIC, Serrano
123, 28006 Madrid, Spain}

\author{P.~Sarriguren}
\affiliation{Instituto de Estructura de la Materia, CSIC, Serrano
123, 28006 Madrid, Spain}

\author{J.~M.~Udias}
\affiliation{Departamento de Fisica Atomica, Molecular y Nuclear,\\
Facultad de Ciencias Fisicas, Universidad Complutense de
Madrid, Madrid E-28040, Spain}

\begin{abstract}
The scaling functions $f(\psi')$ and $F(y)$ from the $\psi'$- and
$y$-scaling analyses of inclusive electron scattering from nuclei
are explored within the coherent density fluctuation model (CDFM).
In addition to the CDFM formulation in which the local density
distribution is used, we introduce a new equivalent formulation of
the CDFM based on the one-body nucleon momentum distribution
(NMD). Special attention is paid to the different ways in which
the excitation energy of the residual system is taken into account
in $y$- and $\psi'$-scaling. Both functions, $f(\psi')$ and
$F(y)$, are calculated using different NMD's and are compared with
the experimental data for a wide range of nuclei. The good
description of the data for $y < 0$ and $\psi' < 0$ (including
$\psi'< -1$) makes it possible to show the sensitivity of the
calculated scaling functions to the peculiarities of the NMD's in
different regions of momenta. It is concluded that the existing
data on the $\psi'$- and $y$-scaling are informative for the NMD's
at momenta not larger than $2.0 \div 2.5$ fm$^{-1}$. The CDFM
allows us to study simultaneously on the same footing the role of
both basic quantities, the momentum and density distributions, for
the description of scaling and superscaling phenomena in nuclei.
\end{abstract}

\pacs{25.30.Fj, 21.60.-n, 21.10.Ft, 24.10.Jv, 21.65.+f }

\maketitle

\section{Introduction}

The inclusive scattering of electrons as weakly interacting probes
from the constituents of a composite nuclear system is a strong
tool in gaining information about the nuclear structure. This
concerns particularly the studies of such basic quantities of the
nuclear ground state as the local density and the momentum
distributions of the nucleons. As known \cite{BS80,JMN85} (see
also \cite{MSCCS91,AHP88,AHP93}) the mean-field approximation
(MFA) is unable to describe simultaneously these two important
nuclear characteristics. This imposes a consistent analysis of the
role of the nucleon-nucleon correlations using theoretical methods
beyond the MFA in the description of the results of the relevant
experiments. It was realized that the nucleon momentum
distribution (NMD), $n(k)$, which is related to both diagonal and
non-diagonal elements of the one-body density matrix is much more
sensitive to the nucleon correlation effects than the density
distribution $\rho(r)$ which is given by its diagonal elements.
Thus it is important to study these two basic characteristics
simultaneously and consistently within the framework of a given
theoretical correlation method analyzing the existing empirical
data. Such a possibility appears in the coherent density
fluctuation model (CDFM) \cite{ANP79+,ANP85,AHP88,AHP93,A+89+}
which is related to the delta-function limit of the generator
coordinate method (see also \cite{AGK+04}). The main aim of the
present work is to apply the CDFM to the description of the
experimental data on the inclusive electron scattering from
nuclei, which showed scaling and superscaling behavior of properly
defined scaling functions, and to gain more information on the NMD
and the density distributions in nuclei.

In the beginning of Section~\ref{s:theo} we will review briefly
the scaling of both first and second kind. Scaling of the first
kind means that in the asymptotic regime of large transfer momenta
$q= |\mathbf{q}|$ and energy $\omega$ a properly defined function
of both of them $F(q,\omega)$ (which is generally the ratio
between the inclusive cross section and the single-nucleon
electromagnetic cross section) becomes a function only of a single
variable, e.g. $y=y(q,\omega)$. This is what is called $y$-scaling
(see,e.g.
\cite{West75,CPS83,CPS91,CW99,CW97,DMD+90,SDM80,PS82,CS96}).
Indeed, for the region $y<0$ and $q>500$ MeV/c this scaling is
quite well obeyed. It has been found that the scaling function is
related to the NMD, and thus some information (though
model-dependent) can be obtained from the $y$-scaling analysis.
Another scaling variable $\psi'$ (related to $y$) and the
corresponding $\psi'$-scaling function $f(\psi')$ were defined and
considered (see, e.g. \cite{AMD+88,BCD+98,DS99,DS99l}) within the
framework of the relativistic Fermi gas (RFG) model. It was found
from the studies of the inclusive scattering cross section data
that $f(\psi')$ shows for $\psi' < 0$ both scaling of the first
kind (independence of $q$) and scaling of the second kind
(independence of the mass number $A$ for a wide range of nuclei
from $^4$He to $^{197}$Au). This is the so called superscaling
\cite{AMD+88}. The extension of the $\psi'$-scaling studies using
the RFG model was given in \cite{MDS02,BCD+04}. Here we would like
to emphasize that, as pointed out in \cite{DS99}, the actual
nuclear dynamical content of the superscaling is more complex than
that provided by the RFG model. For instance, the superscaling
behavior of the experimental data for $f(\psi')$ has been observed
for large negative values of $\psi'$ (up to $\psi' \approx - 2$),
while in the RFG model $f(\psi') = 0$ for $\psi' \leq -1$. This
imposed the consideration of superscaling in theoretical
approaches which go beyond the RFG model, i.e. for realistic
finite nuclear systems. Such a work was performed using the CDFM
in \cite{AGK+04}. The calculations in the model showed a good
quantitative description of the superscaling in finite nuclei for
negative values of $\psi'$, including those smaller than $-1$. We
would like to note that the main ingredient of the CDFM (the
weight function) was expressed and calculated in \cite{AGK+04} on
the basis of experimentally known charge density distributions
$\rho(r)$ for the $^4$He, $^{12}$C, $^{27}$Al, $^{56}$Fe and
$^{197}$Au nuclei. At the same time, however, we started in
\cite{AGK+04} the discussion about the relation of $f(\psi')$ with
the NMD, $n(k)$, showing implicitly how $f(\psi')$ can be
calculated on the basis of $n(k)$. In Ref. \cite{AGK+04} we
indicated an alternative path for defining the weight function of
the CDFM which is built up from a phenomenological or a
theoretical momentum distribution. In the present paper we give
(in Section~\ref{s:theo}) and use (in Section~\ref{s:res}) the
explicit relationship of $f(\psi')$ with $n(k)$ using the basic
scheme of the CDFM and showing also how information about $n(k)$
can be extracted from the $\psi'$-scaling function. We point out
in our theoretical scheme and in our calculations the equivalence
of the cases when $f(\psi')$ is expressed through both density
$\rho(r)$ and momentum distribution $n(k)$. In this way both basic
quantities are used and can be analyzed simultaneously in the
studies of the scaling phenomenon.

Additionally to \cite{AGK+04}, we present calculations of $f(\psi')$
for $q=1560$ MeV/c and the comparison with the experimental data from
\cite{DS99l}.

In the present work we also define the $y$-scaling function $F(y)$
in the CDFM (Section~\ref{s:theo}) and present the comparison with
the experimental data (taken from \cite{CW99,CW97}) of our
calculations of $F(y)$ based on three different NMD's: from the
CDFM, from the $y$-scaling (YS) studies in \cite{CW99,CW97} and
from the parameter-free theoretical approach based on the
light-front dynamics (LFD) method \cite{AGI+02}
(Section~\ref{s:res}). We discuss the sensitivity of the
calculated function $F(y)$ to the peculiarities of the different
NMD's considered.

We also estimate the relationship of $f(\psi')$ with $F(y)$ and show
in Section~\ref{s:res} the condition under which the NMD $n_{CW}(k)$
extracted from the YS analyses \cite{CW99,CW97} can describe the
empirical data on $f(\psi')$.

The consideration of the points mentioned above made it possible to
estimate approximately the region of momenta in $n(k)$ which is
mainly responsible for the description of the $y$- and
$\psi'$-scaling and how it is related to the experimentally studied
regions of the scaling variables $y$ and $\psi'$.

The conclusions of the present work are summarized in
Section~\ref{s:con}.

\section{The theoretical scheme \label{s:theo}}

We start this Section with a brief review of the $y$- and
$\psi'$-scaling analyses in Subsections~\ref{ss:a} and \ref{ss:b},
respectively. An important point that we emphasize is the way in
which the excitation of the residual system is taken into account,
which is different in each case. We discuss the peculiarities of
both approaches that are necessary to take into account for the
development performed in our work within the CDFM
(Subsections~\ref{ss:c} and \ref{ss:d}).

\subsection{Brief review of the $y$-scaling \label{ss:a}}

In this Subsection we will outline the main relationships which
concern the $y$-scaling in the inclusive electron scattering of
high-energy electrons from nuclei (e.g.
\cite{West75,DMD+90,SDM80,PS82,CPS83,CPS91,CW99,CW97}). At large
transfer momentum ($q > 500$ MeV/c) and transfer energy $\omega$,
the scaling function $F(q,\omega)$, which is the cross section of
the inclusive process divided by the elementary probe-constituent
cross section, turns out to be a function of only a single
variable $y=y(q,\omega)$. This is the scaling of the first kind.
The smallest value of the missing momentum $p=|\mathbf{p}|=
|\mathbf{p}_N - \mathbf{q}|$ ($\mathbf{p}_N$ being the momentum of
the outgoing nucleon) at the smallest value of the missing energy
is defined to be $y$ ($-y$) for $\omega$ larger (smaller) than its
value at the quasielastic peak
\begin{equation}\label{2.A.1}
\omega \simeq (q^2 + m_N^2)^{1/2} - m_N,
\end{equation}
$m_N$ being the nucleon mass. The condition for the smallest
missing energy means that the value of the quantity
\begin{equation}\label{2.A.2}
\mathcal{E} (p)=\sqrt{(M_{A-1})^2 + p^2} - \sqrt{(M_{A-1}^0)^2 +
p^2}
\end{equation}
(where $M_{A-1}$ is generally the excited recoiling systems's mass
and $M_{A-1}^0$ is the mass of the system in its ground state) must be
\begin{equation}\label{2.A.3}
\mathcal{E}(p)=0 .
\end{equation}
The quantity $\mathcal{E}(p)$ (\ref{2.A.2}) characterizes the degree
of excitation of the residual system and essentially it is the
missing energy ($E_m$) minus the separation energy ($E_s$). So, at
the condition (\ref{2.A.3}) $E_m = E_s$ .

As shown (e.g.\cite{CPS83,CPS91,CW99,CW97}), for $q > 500$ MeV/c
\begin{equation}\label{2.A.4}
F(q,y)\stackrel{q\rightarrow\infty}\longrightarrow F(y)= f(y)-
B(y) ,
\end{equation}
where
\begin{equation}\label{2.A.5}
f(y)=2\pi\int_{|y|}^{\infty} n(k)k dk
\end{equation}
and $n(k)$ is the conventional NMD function normalized to unity
\begin{equation}\label{2.A.6}
\int d\mathbf{k} n(\mathbf{k})=1 .
\end{equation}

The information on $F(y)$ and, correspondingly, on $f(y)$ can be
used to obtain $n(k)$ by:
\begin{equation}\label{2.A.7}
n(k)= \left. - \frac{1}{2\pi y} \frac{df(y)}{dy} \right |_{|y|=k}.
\end{equation}
In Eq. (\ref{2.A.4}) $B(y)$ is the binding correction which is
related to the part of the spectral function generated by
ground-state correlations and the excitations of the residual
system (when $M_{A-1} > M^0_{A-1}$ and, correspondingly,
$\mathcal{E}(p)
> 0$ ).

The problem of the correct account for the binding correction is a
longstanding one. Only when the excitation energy of the residual
system is equal to zero (as in the case of the deuteron) $B=0$ and
then $F(y)= f(y)$. Generally, however, the final system of $A-1$
nucleons can be left in all possible excited states. Then $B(y)
\neq 0$ and $F(y) \neq f(y)$.

In \cite{CW99} a new $y$-scaling variable ( $y_{CW}$) was
introduced on the basis of a realistic nuclear spectral function
as provided by few- and many-body calculations
\cite{M+80C+,BFF92}. The use of $y_{CW}$ leads to $B(y_{CW}) = 0$
and, consequently, to $F(y_{CW}) = f(y_{CW})$. The latter is
important because in this case it becomes possible to obtain
information on the NMD directly (using Eq.~(\ref{2.A.7})) without
introducing theoretical binding correction $B(y)$. In this
consideration the removal energy (whose effects are a source of
scaling violation, the other source being the final-state
interactions) is taken into account in the definition of the
scaling variable. So, the binding corrections are incorporated
into the definition of $y_{CW}$.

The analysis of empirical data on inclusive electron scattering
from nuclei (with $A \leq 56$) showed \cite{CW99,CW97} that the
following form of $f(y)$ gives a very good agreement with the
data:
\begin{equation}\label{2.A.8}
f(y)= \frac{C_1\exp(-a^2y^2)}{\alpha^2+y^2}+ C_2\exp(-b|y|) \left(
1+ \frac{by}{1+ y^2/\alpha^2} \right),
\end{equation}
where the first term describes the small $y$-behavior and the
second term dominates large $y$. From Eq.~(\ref{2.A.7}) one can
obtain the following form of $n(k)$:
\begin{equation}\label{2.A.9}
n(k)= n_{MFA}(k)+ n_{corr}(k),
\end{equation}
where the mean-field part $n_{MFA}(k)$ of the NMD (for $k \lesssim
2$ fm$^{-1}$) is:
\begin{equation}\label{2.A.10}
n_{MFA}(k)= \frac{C_1}{\pi} \left[ 1+ a^2(\alpha^2 + k^2)\right]
\frac{\exp(-a^2 k^2)}{(\alpha^2 + k^2)^2},
\end{equation}
while the high-momentum components of $n(k)$ which contain nucleon
correlation effects are given by:
\begin{equation}\label{2.A.11}
n_{corr}(k)= \frac{C_2 b\exp(-bk)}{2\pi(1+ k^2/\alpha^2)} \left[
b+ \frac{k}{\alpha^2} \left( 3+ bk+
\frac{k^2}{\alpha^2}\right)\right] .
\end{equation}
Further in our work we will use the information about the $n(k)$
from the $y$-scaling analysis [Eqs.~(\ref{2.A.9})-(\ref{2.A.11})].
The values of the parameters \cite{CW99,CW97}, e.g. in the case of
interest for the $^{56}$Fe nucleus are: $b=1.1838$ fm, $C_1=0.30$
fm$^{-1}$, $C_2=0.11838$ fm, $\alpha =0.710$ fm$^{-1}$ and
$a=0.908$ fm.

\subsection{The $\psi'$-scaling variable and the $\psi'$-scaling
function in the relativistic Fermi gas model. Relation between the
$y$- and $\psi'$-scaling variables \label{ss:b}}

In this Subsection we will review briefly the scaling in the
framework of the RFG model \cite{AMD+88,BCD+98,DS99,DS99l}. This
will be necessary for our consideration of the scaling in the
present work within the CDFM in the next Subsections~\ref{ss:c}
and \ref{ss:d}. The $y$-scaling variable in the RFG has the form
\begin{equation}\label{2.B.1}
y_{RFG}= m_N \left( \lambda \sqrt{1+ \frac{1}{\tau}} -
\kappa\right) ,
\end{equation}
where
\begin{equation}\label{2.B.2}
\kappa \equiv q/2m_N,\;\;\; \lambda \equiv \omega/2m_N,\;\;\; \tau
\equiv |Q^2|/4m_N^2= \kappa^2 - \lambda^2
\end{equation}
are the dimensionless versions of $q$, $\omega$ and the squared
four-momentum $|Q^2|$. In \cite{AMD+88,BCD+98,DS99,DS99l} a new
scaling variable $\psi$ was introduced by
\begin{equation}\label{2.B.3}
\psi= \frac{1}{\sqrt{\xi_F}} \frac{\lambda - \tau}{\sqrt{(1+
\lambda)\tau+ \kappa\sqrt{\tau(1+ \tau)}}} ,
\end{equation}
where
\begin{equation}\label{2.B.4}
\xi_F= \sqrt{1+ \eta_F^2}- 1 \;\;\; \textrm{and} \;\;\;
\eta_F=k_F/m
\end{equation}
are the dimensionless Fermi kinetic energy and Fermi momentum,
respectively.

In order to include, at least partially, the missing energy
dependence in the scaling variable, a shift of the energy $\omega$ is
introduced in the RFG \cite{DS99}:
\begin{equation}\label{2.B.5}
\omega'\equiv \omega - E_{shift},
\end{equation}
where $E_{shift}$ is chosen empirically (in practice it is from 15
to 25 MeV) and thus can take values other than the separation
energy $E_s$. The corresponding $\lambda$ and $\tau$ become
\begin{equation}\label{2.B.6}
\lambda'\equiv\omega'/2m_N, \;\;\; \tau'\equiv\kappa^2-
{\lambda'}^2 .
\end{equation}
This procedure aims to account for the effects of both binding in the
initial state and interaction strength in the final state. It is
shown in \cite{DS99} that the corresponding new version of the
$\psi$-scaling variable ($\psi'$) has the following relation to the
$y$-scaling variable:
\begin{eqnarray} \label{2.B.7}
\psi' \equiv \psi [\lambda \rightarrow \lambda'] &=& \frac{y_{\infty}
(\widetilde{\lambda}=\lambda')}{k_F} \left( 1+ \sqrt{1+
\frac{1}{4\kappa^2}} \frac{1}{2}\eta_F \frac{y_{\infty}
(\widetilde{\lambda}=\lambda')}{k_F}\right) + \mathcal{O}\left[
\eta_F^2 \right] ,\\
\nonumber&& \widetilde{\lambda} \equiv
\frac{\widetilde{\omega}}{2m_N}= \frac{\omega - E_s}{2m_N} .
\end{eqnarray}
In (\ref{2.B.7}) $y_{\infty}$ is the $y$-scaling variable in the
limit where $M^0_{A-1} \longrightarrow \infty$. $k_F$ is the Fermi
momentum which is a free parameter in the RFG model, taking values
from 1.115 fm$^{-1}$ for $^{12}$C to 1.216 fm$^{-1}$ for
$^{197}$Au \cite{DS99}. As shown in \cite{DS99}, Eq.~(\ref{2.B.7})
contains an important average dependence on the quantity
$\mathcal{E}(p)$ (\ref{2.A.2}) (i.e. on the missing energy) which
is reflected in the quadratic dependence of $\psi'$ on the
$y$-scaling variable.

Finally, in \cite{DS99,DS99l} a dimensionless scaling function is
introduced within the RFG model
\begin{equation}\label{2.B.8}
f_{RFG}(\psi')=k_F F_{RFG}(\psi') .
\end{equation}
The careful analysis of the experimental data on inclusive
electron scattering \cite{DS99,DS99l} shows that the RFG model
contains scaling of the first kind ($f$ or $F$ are not dependent
on $q$ at high-momentum transfer and depend only on $\psi'$) but
also that $f(\psi')$ is independent of $k_F$ to leading order in
$\eta_F^2$, thus showing no dependence on the mass number $A$
(scaling of the second kind). In the RFG both kinds of scaling
occur and this phenomenon is called superscaling.

To finalize this Subsection we give the analytical form of
$f_{RFG}$ obtained in \cite{AMD+88,BCD+98,DS99,DS99l} which will
be used in this work:
\begin{equation}\label{2.B.9}
f_{RFG}(\psi')= \frac{3}{4}(1-{\psi'}^2)\Theta
(1-{\psi'}^2)\frac{1}{\eta_F^2} \left[ \eta_F^2 +
{\psi'}^2\left(2+ \eta_F^2- 2\sqrt{1+ \eta_F^2}\right) \right] .
\end{equation}
Here we note that due to the $\Theta$-function in (\ref{2.B.9}),
the function $f(\psi')$ is equal to zero at $\psi' \leq -1$ and
$\psi' \geq 1$. As can be seen in Fig.~1 of Ref. \cite{AGK+04},
this is not in accordance with the experimental data and justifies
the attempt in \cite{AGK+04}, as well as the development made in
the present work, to consider the superscaling in realistic
systems beyond the RFG model.

\subsection{Theoretical scheme of the CDFM and the $\psi'$-scaling
function in the model \label{ss:c}}

The CDFM was suggested and developed in
\cite{ANP79+,ANP85,AHP88,AHP93,A+89+}.  It was deduced from the
delta-function limit of the generator coordinate method \cite{GW57}.
The model was applied to the study of the superscaling phenomenon in
\cite{AGK+04}. In this Subsection we continue the development of the
model aiming its applications to the studies of the NMD from the
analyses of the $y$- and $\psi'$-scaling in inclusive electron
scattering from nuclei. We will start with the expressions of the
Wigner distribution function (WDF) in the CDFM
$W(\mathbf{r},\mathbf{p})$ (e.g.\cite{AHP88,AHP93}). They are based
on two representations of the WDF for a piece of nuclear matter which
contains all $A$ nucleons distributed homogeneously in a sphere with
radius $R$, with density
\begin{equation}\label{2.C.1}
\rho_{0}(R)=\frac{3A}{4\pi R^{3}}
\end{equation}
and Fermi momentum
\begin{equation}
\overline{k}_F= k_{F}(R)=\left(\frac{3\pi^{2}}{2}\rho_{0}(R)\right
)^{1/3}\equiv \frac{\alpha}{R}, \;\;\; \alpha=\left(\frac{9\pi
A}{8}\right )^{1/3}\simeq 1.52A^{1/3} . \label{2.C.2}
\end{equation}
The first form of the WDF is:
\begin{equation}
W_{R}(\mathbf{r},\mathbf{p})=\frac{4}{(2\pi)^{3}}\Theta
(R-|\mathbf{r}|)\Theta (k_{F}(R)-|\mathbf{p}|). \label{2.C.3}
\end{equation}
The second form of the WDF for such a piece of nuclear matter can
be written as
\begin{equation}
W_{\overline{k}_F}(\mathbf{r},\mathbf{p})=\frac{4}{(2\pi)^{3}}\Theta
(\overline{k}_F-|\mathbf{p}|)\Theta (\frac{\alpha}{\overline{k}_F}
-|\mathbf{p}|). \label{2.C.4}
\end{equation}
In the CDFM the WDF, as well as the corresponding one-body density
matrix (ODM) can be written as superpositions of WDF's (ODM's) from
Eqs.~(\ref{2.C.3}) and (\ref{2.C.4}) in coordinate and momentum
space, respectively:
\begin{equation}
W(\mathbf{r},\mathbf{p})=\int_{0}^{\infty}dR|F(R)|^2
W_R(\mathbf{r},\mathbf{p})=
\frac{4}{(2\pi)^{3}}\int_{0}^{\infty}dR|F(R)|^2 \Theta
(R-|\mathbf{r}|)\Theta (k_{F}(R)-|\mathbf{p}|) \label{2.C.5}
\end{equation}
and
\begin{equation}
W(\mathbf{r},\mathbf{p})= \int_{0}^{\infty}
d\overline{k}_F|G(\overline{k}_F)|^2
W_{\overline{k}_F}(\mathbf{r},\mathbf{p})= \frac{4}{(2\pi)^{3}}
\int_{0}^{\infty}d\overline{k}_F|G(\overline{k}_F)|^2 \Theta
(\overline{k}_F-|\mathbf{p}|)\Theta
(\frac{\alpha}{\overline{k}_F}-|\mathbf{r}|) . \label{2.C.6}
\end{equation}
The relationship between both $|F|^2$ and $|G|^2$ functions is:
\begin{equation}\label{2.C.7}
|G(\overline{k}_F)|^2= \frac{\alpha}{{\overline{k}_F}^2}
|F(\frac{\alpha}{\overline{k}_F})|^2 .
\end{equation}
Using the basic relationships of the density and momentum
distributions with the WDF:
\begin{equation}\label{2.C.8}
\rho(\mathbf{r})=\int d\mathbf{p}W(\mathbf{r},\mathbf{p}),
\end{equation}
\begin{equation}\label{2.C.9}
n(\mathbf{p})=\int d\mathbf{r}W(\mathbf{r},\mathbf{p}) ,
\end{equation}
one can obtain the corresponding expressions for $\rho(r)$ and
$n(p)$ using the WDF from Eq.~(\ref{2.C.5}):
\begin{equation}\label{2.C.10}
\rho(\mathbf{r})=\int_{0}^{\infty}dR|F(R)|^{2} \frac{3A}{4\pi R^3}
\Theta (R-|\mathbf{r}|) ,
\end{equation}
\begin{equation}\label{2.C.11}
n(\mathbf{p})=\frac{2}{3\pi^{2}}\int_{0}^{\alpha/p} dR|F(R)|^{2}
R^{3},
\end{equation}
and, equivalently, using the WDF from Eq.~(\ref{2.C.6}):
\begin{equation}\label{2.C.12}
\rho(\mathbf{r})= \frac{2}{3\pi^{2}} \int_{0}^{\infty}
d\overline{k}_F |G(\overline{k}_F)|^2 \Theta (\frac{\alpha}{r}-
\overline{k}_F) {\overline{k}_F}^3,
\end{equation}
\begin{equation}\label{2.C.13}
n(\mathbf{p})= \int_{0}^{\infty} d\overline{k}_F
|G(\overline{k}_F)|^2 \frac{3A}{4\pi {\overline{k}_F}^3} \Theta
(\overline{k}_F-|\mathbf{p}|) ,
\end{equation}
both normalized to the mass number
\begin{equation}\label{2.C.14}
\int \rho(\mathbf{r})d\mathbf{r}=A, \;\;\; \int
n(\mathbf{k})d\mathbf{k}=A
\end{equation}
when both weight functions are normalized to unity:
\begin{equation}\label{2.C.15}
\int_{0}^{\infty}dR|F(R)|^{2} =1, \;\;\; \int_{0}^{\infty}
d\overline{k}_F |G(\overline{k}_F)|^2=1 .
\end{equation}
One can see from Eqs.~(\ref{2.C.10}), (\ref{2.C.11}) and
(\ref{2.C.12}), (\ref{2.C.13}) the symmetry of the expressions for
$\rho(r)$ and $n(p)$ as integrals in the coordinate and momentum
space.

A convenient approach to obtain the weight functions $F(R)$ and
$G(\overline{k}_F)$ is to use a known (experimental or
theoretical) density distribution $\rho(r)$ and/or the momentum
distribution $n(p)$ for a given nucleus. For $|F(R)|^2$ one can
obtain from Eqs.~(\ref{2.C.10}) and (\ref{2.C.11}):
\begin{equation}
|F(R)|^{2}=-\frac{1}{\rho_{0}(R)} \left. \frac{d\rho(r)}{dr}\right
|_{r=R} \label{2.C.16}
\end{equation}
(at $d\rho/dr\leq 0$) and
\begin{equation}
|F(R)|^{2}=-\frac{3\pi^2}{2} \frac{\alpha}{R^5}\left.
\frac{dn(p)}{dp}\right |_{p=\alpha/R} \label{2.C.17}
\end{equation}
(at $dn/dp \leq 0$).

The expressions for $|G(\overline{k}_F)|^2$ can be obtained from
Eqs.~(\ref{2.C.12}) and (\ref{2.C.13}):
\begin{equation}
|G(\overline{k}_F)|^2=-\frac{3\pi^2}{2}
\frac{\alpha}{{\overline{k}_F}^5}\left. \frac{d\rho(r)}{dr}\right
|_{r=\alpha/\overline{k}_F} \label{2.C.18}
\end{equation}
(at $d\rho/dr\leq 0$) and
\begin{equation}
|G(\overline{k}_F)|^2=- \frac{1}{n_0(\overline{k}_F)}\left.
\frac{dn(p)}{dp}\right |_{p=\overline{k}_F} \label{2.C.19}
\end{equation}
(at $dn/dp \leq 0$) with
\begin{equation}\label{2.C.20}
n_0(\overline{k}_F)= \frac{3A}{4\pi {\overline{k}_F}^3} .
\end{equation}

In order to introduce the scaling function within the CDFM, we
assume that the scaling function for a finite nucleus $f(\psi')$
can be defined and obtained by means of the weight function
$|F(R)|^2$ (and $|G(\overline{k}_F)|^2$) weighting the scaling
function for the RFG model depending on the scaling variable
$\psi'_{R}$ ($f_{RFG}(\psi'=\psi'_{R})$, Eq.~(\ref{2.B.9})),
corresponding to a given density $\rho_0(R)$ (\ref{2.C.1}) and
Fermi momentum $k_F(R)$ (\ref{2.C.2}) \cite{AGK+04} (or
corresponding to a given density in the momentum space
$n_0(\overline{k}_F)$ (\ref{2.C.20})).

One can write the scaling variable $\psi'_R$ in the form
\cite{AGK+04}:
\begin{equation}
\psi'_{R}(y)=\frac{p(y)}{k_F(R)}=\frac{p(y)R}{\alpha},
\label{2.C.21}
\end{equation}
where
\begin{equation}\label{2.C.22}
p(y)=\left\{
\begin{array}{ccl}
y(1+cy), & \;\;\; &y\geq 0 \\
-|y|(1-c|y|), & &y\leq 0, |y|\leq 1/2c \\
\end{array}
\right.
\end{equation}
with
\begin{equation}
c\equiv \frac{1}{2m_{N}}\sqrt{1+\frac{1}{4\kappa^{2}}}.
\label{2.C.23}
\end{equation}
Also a more convenient notation can be used:
\begin{equation}
\psi'_{R}(y)= \frac{k_{F}}{k_{F}(R)} \frac{p(y)}{k_{F}}=
\frac{k_{F}}{k_{F}(R)} \psi'. \label{2.C.24}
\end{equation}
Using the $\Theta$-function in Eq.~(\ref{2.B.9}) the scaling
function for a finite nucleus can be defined by the following
expressions:
\begin{equation}\label{2.C.25}
f(\psi')= \int_{0}^{\alpha/(k_{F}|\psi'|)}dR |F(R)|^{2}
f_{RFG}(R,\psi')
\end{equation}
with
\begin{eqnarray}
f_{RFG}(R,\psi')& =& \displaystyle \frac{3}{4} \left[ 1-\left(
\frac{k_FR|\psi'|}{\alpha} \right)^{2}\right] \left\{ 1+ \left(
\frac{Rm_N}{\alpha}\right)^2 \left( \frac{k_FR|\psi'|}{\alpha}
\right)^2 \right. \nonumber\\
&& \times \displaystyle \left. \left[2+ \left( \frac{\alpha}{Rm_N}
\right)^2- 2\sqrt{1+ \left( \frac{\alpha}{Rm_N} \right)^2}\right]
\right\} \label{2.C.26}
\end{eqnarray}
and also, equivalently, by
\begin{equation}\label{2.C.27}
f(\psi')= \int_{k_{F}|\psi'|}^{\infty} d\overline{k}_F
|G(\overline{k}_F)|^2 f_{RFG}(\overline{k}_F, \psi')
\end{equation}
with
\begin{eqnarray}
f_{RFG}(\overline{k}_F,\psi')& =& \displaystyle \frac{3}{4} \left[
1-\left( \frac{k_F|\psi'|}{\overline{k}_F} \right)^2 \right]
\left\{ 1+ \left( \frac{m_N}{\overline{k}_F} \right)^2 \left(
\frac{k_F|\psi'|}{\overline{k}_F} \right)^2 \right. \nonumber\\
&& \times \displaystyle \left. \left[2+ \left(
\frac{\overline{k}_F}{m_N} \right)^2- 2\sqrt{1+ \left(
\frac{\overline{k}_F}{m_N} \right)^2}\right] \right\} .
\label{2.C.28}
\end{eqnarray}
In this way in the CDFM the scaling function $f(\psi')$ is an
infinite superposition of the RFG scaling functions
$f_{RFG}(R,\psi')$ (or $f_{RFG}(\overline{k}_F,\psi')$).

In Eqs.~(\ref{2.C.25})-(\ref{2.C.28}) the momentum $k_F$ is not a
free fitting parameter for different nuclei, as in the RFG model,
but can be calculated consistently in the CDFM for each nucleus
(see (\ref{2.C.16})-(\ref{2.C.19})) using the expression
\begin{equation} \label{2.C.29}
k_F= \int_{0}^{\infty} dR k_{F}(R)|F(R)|^2= \alpha
\int_{0}^{\infty} dR \frac{1}{R}|F(R)|^{2}=
\frac{4\pi(9\pi)^{1/3}}{3A^{2/3}} \int_{0}^{\infty} dR \rho(R) R
\end{equation}
when the condition
\begin{equation}\label{2.C.30}
\lim_{R\rightarrow \infty} \left[ \rho(R)R^2 \right]=0
\end{equation}
is fulfilled and, equivalently,
\begin{equation}\label{2.C.31}
k_F= \frac{16\pi}{3A} \int_0^\infty d\overline{k}_F
n(\overline{k}_F ) {\overline{k}_F}^3,
\end{equation}
when the condition
\begin{equation}\label{2.C.32}
\lim_{\overline{k}_F\rightarrow \infty}\left[
n(\overline{k}_F){\overline{k}_F}^4 \right]=0
\end{equation}
is fulfilled. Generally, Eqs.~(\ref{2.C.30}) and (\ref{2.C.32})
are fulfilled, so the Eqs.~(\ref{2.C.29}) and (\ref{2.C.31}) can
be used to calculate $k_F$ in most of the cases of interest.

The integration in (\ref{2.C.25}) and (\ref{2.C.27}), using
Eqs.~(\ref{2.C.16})-(\ref{2.C.19}), leads to the following
expressions for $f(\psi')$:
\begin{equation}\label{2.C.33}
f(\psi')= \frac{4\pi}{A}\int_{0}^{\alpha/(k_{F}|\psi'|)}dR \rho(R)
\left[ R^2 f_{RFG}(R,\psi')+ \frac{R^3}{3} \frac{\partial
f_{RFG}(R,\psi')}{\partial R} \right]
\end{equation}
and
\begin{equation}\label{2.C.34}
f(\psi')= \frac{4\pi}{A} \int_{k_{F}|\psi'|}^{\infty}
d\overline{k}_F n(\overline{k}_F) \left[ {\overline{k}_F}^2
f_{RFG}(\overline{k}_F, \psi')+ \frac{{\overline{k}_F}^3}{3}
\frac{\partial f_{RFG}(\overline{k}_F,\psi')}{\partial
\overline{k}_F} \right]
\end{equation}
the latter at
\begin{equation}\label{2.C.35}
\lim_{\overline{k}_F\rightarrow \infty}\left[
n(\overline{k}_F){\overline{k}_F}^3 \right]=0 ,
\end{equation}
where the functions $f_{RFG}(R,\psi')$ and
$f_{RFG}(\overline{k}_F, \psi')$ are given by Eqs.~(\ref{2.C.26})
and (\ref{2.C.28}), respectively. We emphasize the symmetry in
both Eqs.~(\ref{2.C.33}) and (\ref{2.C.34}). We also note that the
CDFM scaling function $f(\psi')$ is symmetric at the change of
$\psi'$ to $-\psi'$.

The scaling function $f(\psi')$ can be calculated using
Eqs.~(\ref{2.C.33}) and (\ref{2.C.34}) by means of: i) its
relationship to the density distribution $\rho(r)$, and ii) from
the relationship to the NMD $n(p)$. Both quantities ($\rho$ and
$n$) can be taken from empirical data or from theoretical
calculations. In the CDFM they are consistently related because
they are based on the WDF of the model (Eqs.~(\ref{2.C.5}) and
(\ref{2.C.6})). Using experimentally known density distributions
$\rho(r)$ for a given nucleus one can calculate the weight
functions $|F|^2$ (Eq.~(\ref{2.C.16})) or $|G|^2$
(Eq.~(\ref{2.C.18})) and by means of them to calculate $n(p)$ in
the CDFM (by Eqs.~(\ref{2.C.11}) or (\ref{2.C.13}), respectively).

From Eq.~(\ref{2.C.34}) one can estimate the possibility to obtain
information about the NMD from the empirical data on the scaling
function $f(\psi')$. If we keep only the main term of the function
$f_{RFG}(\overline{k}_F, \psi')$
\begin{equation}\label{2.C.36}
f_{RFG}(\overline{k}_F, \psi')\simeq \frac{3}{4} \left( 1-
\frac{\left( k_F |\psi'| \right)^2}{{\overline{k}_F}^2} \right)
\end{equation}
and its derivative
\begin{equation}\label{2.C.37}
\frac{\partial f_{RFG}(\overline{k}_F,\psi')}{\partial
\overline{k}_F}\simeq \frac{3}{2} \frac{\left( k_F |\psi'|
\right)^2}{{\overline{k}_F}^3} ,
\end{equation}
then:
\begin{equation}\label{2.C.38}
f(\psi')\simeq 3\pi \int_{k_F|\psi'|}^{\infty} d\overline{k}_F
n(\overline{k}_F) {\overline{k}_F}^2 \left[ 1 -
\frac{1}{3}\frac{\left( k_F |\psi'| \right)^2}{{\overline{k}_F}^2}
\right] .
\end{equation}
In Eq.~(\ref{2.C.38})
\begin{equation}\label{2.C.39}
\int n(\overline{\mathbf{k}}_F) d\overline{\mathbf{k}}_F=1.
\end{equation}
Neglecting the second term in the bracket in (\ref{2.C.38})
(because $\displaystyle \frac{1}{3}\frac{\left( k_F |\psi'|
\right)^2}{{\overline{k}_F}^2} \ll 1$) one obtains:
\begin{equation}\label{2.C.40}
f(\psi')\simeq 3\pi \int_{k_F|\psi'|}^{\infty} d\overline{k}_F
n(\overline{k}_F) {\overline{k}_F}^2.
\end{equation}
Taking the derivative on $|\psi'|$ from both sides of
Eq.~(\ref{2.C.40}) leads to:
\begin{equation}\label{2.C.41}
n(p)= - \left. \frac{1}{3\pi p^2 k_F} \frac{\partial
f(\psi')}{\partial (|\psi'|)} \right|_{|\psi'|=p/k_F} .
\end{equation}
Eq.~(\ref{2.C.41}) can give approximately information on the
NMD $n(p)$. If one keeps the second term in the bracket under the
integral in (\ref{2.C.38}), then more complicate equation results:
\begin{equation}\label{2.C.42}
\left. \frac{\partial f(\psi')}{\partial (k_F |\psi'|)}
\right|_{k_F|\psi'|=p}= -2\pi p^2 n(p) -2\pi p\int_p^\infty dk'
n(k') .
\end{equation}

\subsection{$y$-scaling function in the CDFM and the relationship
between the $y$- and $\psi'$-scaling functions in the model
\label{ss:d}}

In order to define the $y$-scaling function $F(y)$ (with
dimensions) in the CDFM and to establish the relationship between
the latter and the dimensionless scaling function $f(\psi')$
(Eqs.~(\ref{2.C.33}) and (\ref{2.C.34})) we start with the
expression which relates both functions in the RFG model
\cite{DS99,DS99l}:
\begin{equation}\label{2.D.1}
F_{RFG}(y)=
\frac{f_{RFG}({\overline{k}_F},\psi'(y))}{{\overline{k}_F}}.
\end{equation}
In analogy with the definition of $f(\psi')$ in the CDFM, we
introduce the function $F(y)$ in a finite system as a
superposition of RFG $y$-scaling functions $F_{RFG}(y)$
(\ref{2.D.1}):
\begin{equation}\label{2.D.2}
F(y)= \int_{k_{F}|\psi'|= |p(y)|}^{\infty} d\overline{k}_F
|G(\overline{k}_F)|^2 \frac{f_{RFG}(k_{F}|\psi'|= |p(y)|,
\overline{k}_F)}{\overline{k}_F} ,
\end{equation}
where the function $f_{RFG}$ has the form (\ref{2.C.28}) and
$k_F|\psi'|= |p(y)|$ with $p(y)$ given by Eq.~(\ref{2.C.22}).
Using Eq.~(\ref{2.C.19}) for $|G(\overline{k}_F)|^2$, we obtain
from Eq.~(\ref{2.D.2}):
\begin{equation}\label{2.D.3}
F(y)= \frac{4\pi}{3} \int_{k_{F}|\psi'|= |p(y)|}^{\infty}
d\overline{k}_F n(\overline{k}_F) \left\{ 2k f_{RFG}(|p(y)|,
\overline{k}_F)+ {\overline{k}_F}^2 \frac{\partial f_{RFG}(|p(y)|,
\overline{k}_F)}{\partial \overline{k}_F} \right\} .
\end{equation}
Keeping in Eq.~(\ref{2.D.3}) the main terms of the function
$f_{RFG}$ [Eq.~(\ref{2.C.36})] and of its derivative $\partial
f_{RFG}/\partial \overline{k}_F$ [Eq.~(\ref{2.C.37})], one obtains
the scaling function $F(y)$ in the form:
\begin{equation}\label{2.D.4}
F(y)\simeq 2\pi \int_{|p(y)|}^{\infty} dk k n(k) .
\end{equation}
In Eqs.~(\ref{2.D.3}) and (\ref{2.D.4}) the normalization of
$n(k)$ is:
\begin{equation}\label{2.D.5}
\int n(\mathbf{k}) d\mathbf{k}=1 .
\end{equation}
For the cases of interest when $y\leq 0$ ($|y|\leq 1/(2c)$):
\begin{equation}\label{2.D.6}
F(y)\simeq 2\pi \int_{|y|(1- c|y|)}^\infty dk \, k n(k) .
\end{equation}

If the $\psi'$-scaling variable is not a quadratic function of $y$
as in \cite{DS99,DS99l} (see Eq.~(\ref{2.B.7})) but it is a linear
one (i.e. $k_F|\psi'|= |y|$), then
\begin{equation}\label{2.D.7}
F(y)\simeq 2\pi \int_{|y|}^\infty dk \, k n(k) .
\end{equation}
Eq.~(\ref{2.D.7}) gives the known $y$-scaling function
\cite{West75,CPS83,CPS91,CW99,CW97} and its relationship with the
NMD $n(k)$. It can be seen from (\ref{2.D.6}) that the use of a
more complicated (quadratic) dependence of $\psi'$ on $y$
\cite{DS99,DS99l} leads to a more complicated $y$-scaling function
in the CDFM and its relationship with the NMD.

To finalize this section we elaborate somewhat more on
Eq.~(\ref{2.C.40}), considering the relationship between the
scaling variables $y$ and $\psi'$ and between the corresponding
scaling functions. Eq.~(\ref{2.C.40}) for the $\psi'$-scaling
function $f(\psi')$ can be rewritten in the following approximate
form:
\begin{equation}\label{2.D.8}
f(\psi')\simeq 3\pi \int_{k_F|\psi'|}^{\infty} d\overline{k}_F
n(\overline{k}_F) {\overline{k}_F}^2 \simeq \frac{3}{2} k_{av}
2\pi \int_{k_F|\psi'|}^{\infty} d\overline{k}_F n(\overline{k}_F)
{\overline{k}_F} .
\end{equation}
If we admit as in the case of the $y$-scaling
\cite{West75,CPS83,CPS91,CW99,CW97}
\begin{equation}\label{2.D.9}
k_F|\psi'|= |y|,
\end{equation}
then
\begin{equation}\label{2.D.10}
f(\psi')= \frac{3}{2} k_{av} F(y),
\end{equation}
where $F(y)$ is the $y$-scaling function (\ref{2.D.7}) and
$k_{av}$ can be estimated as in \cite{CW99}:
\begin{equation}\label{2.D.11}
k_{av}\simeq \left\langle \frac{1}{k} \right\rangle^{-1}, \;\;\;
\text{where} \;\;\; \left\langle \frac{1}{k} \right\rangle = \int
d\mathbf{k} \frac{n(k)}{k}.
\end{equation}
In the case of the $\psi'$-scaling variable (for $y\leq 0$)
\begin{equation}\label{2.D.12}
k_F|\psi'|= |y|(1-c|y|)
\end{equation}
and we can replace approximately the lower limit of the
integration by $|y|$ which is, however, the solution of
Eq.~(\ref{2.D.12}):
\begin{equation}\label{2.D.13}
f(\psi')\simeq 3\pi \int_{|y|}^{\infty} d\overline{k}_F \,
n(\overline{k}_F) {\overline{k}_F}^2 , \;\;\; \text{where} \;\;\;
|y|=\frac{1}{2c}\left( 1- \sqrt{1-4ck_F |\psi'|}\right) .
\end{equation}

\section{results of calculations and discussion \label{s:res}}

\begin{figure}[htb]
\includegraphics[width=12cm]{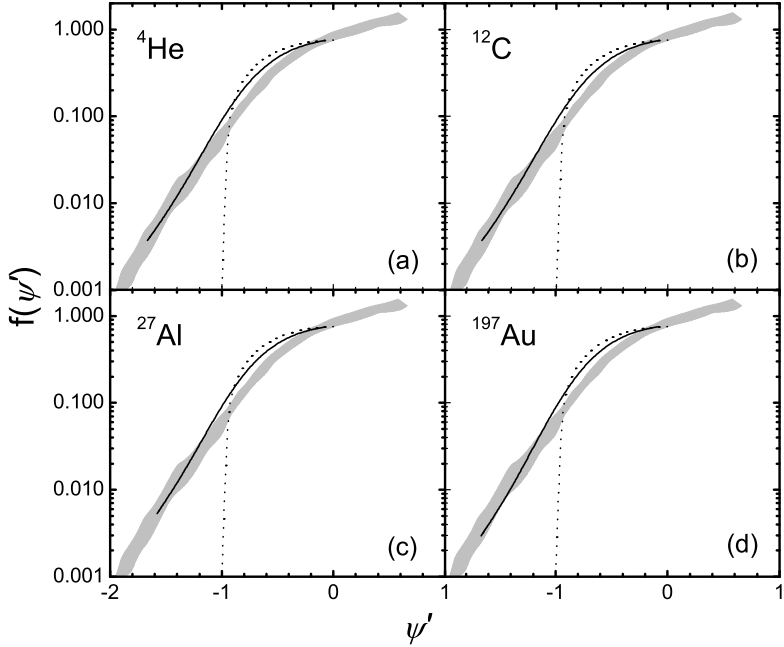}\\
\caption{Scaling function $f(\psi')$ in the CDFM (solid line) at
$q=1560$ MeV/c for $^4$He, $^{12}$C, $^{27}$Al, and $^{197}$Au.
The results are obtained using Eqs.~(\protect\ref{2.C.33}),
(\protect\ref{2.C.29}) and equivalently by Eqs.
(\protect\ref{2.C.34}), (\protect\ref{2.C.31}) with $n(p)$ from
the CDFM [Eq.~(\protect\ref{2.C.11})]. The experimental data from
\protect\cite{DS99l} are given by the shaded area. The RFG result
[Eq.~(\protect\ref{2.B.9})] is shown by dotted line.
\label{fig1}}
\end{figure}

We begin this Section with calculations of the scaling function
$f(\psi')$ using Eqs.~(\ref{2.C.33}) and (\ref{2.C.34}) for
different nuclei within the CDFM for the transfer momentum
$q=1560$ MeV/c. The results for $^4$He, $^{12}$C, $^{27}$Al and
$^{197}$Au are presented in Fig.~\ref{fig1} and are compared with
the experimental data from \cite{DS99l} and with the predictions
of the RFG model [Eq.~(\ref{2.B.9})] with values of $k_F$ from
\cite{DS99,DS99l}. Our calculations are performed in addition to
those for $q=1000$ and 1650 MeV/c in \cite{AGK+04} which were
compared with the data from \cite{DS99}.

Following the main aim of our work, namely to extract reliable
information on the NMD from the scaling function, we will consider
in detail the consecutive steps of the calculations of $f(\psi')$
in connection with $n(k)$. We note firstly that $f(\psi')$ is
calculated in the CDFM using Eq.~(\ref{2.C.33}) where the density
distribution $\rho(r)$ is taken to be of a Fermi-type with
parameter values obtained from the experimental data on elastic
electron scattering from nuclei and muonic atoms. For $^4$He and
$^{12}$C we used a symmetrized Fermi-type density distribution
\cite{BKL+98} with half-radius ($R_{1/2}$) and diffuseness ($b$)
parameters: $R_{1/2} = 1.710$ fm, $b = 0.290$ fm for $^4$He and
$R_{1/2}= 2.470$ fm, $b= 0.420$ fm for $^{12}$C. These values of
the parameters lead to charge rms radii equal to 1.710 fm for
$^4$He and 2.47 fm for $^{12}$C which coincide with the
experimental ones \cite{VJV87}. For the $^{27}$Al nucleus the
values of $R_{1/2} =3.070$ fm and $b= 0.519$ fm are taken from
\cite{VJV87}. For $^{197}$Au the parameter values are $R_{1/2} =
6.419$ fm \cite{PP03} and $b = 1.0$ fm \cite{AGK+04}. The
necessity to use the latter value of $b$ for $^{197}$Au instead of
$b = 0.449$ fm \cite{PP03} was discussed in \cite{AGK+04}. This is
an ad-hoc procedure in order to obtain high-momentum components of
the NMD $n(k)$ (using (\ref{2.C.11}) and (\ref{2.C.16})) which are
similar to those in light and medium nuclei. This was necessary to
be done due to the particular $A$-dependence of $n(k)$ in the CDFM
resulting in lower tails of $n(k)$ at $k>2$ fm$^{-1}$ for the
heaviest nuclei which has to be improved. Also for $^{56}$Fe a
better agreement with $\psi'$ scaling data at $q=1000$ MeV/c is
obtained for $b = 0.7$ fm instead of $b = 0.558$ fm \cite{VJV87}
and this result will be shown later on.

As can be seen, the CDFM results for the scaling function
$f(\psi')$ agree well with the experimental data taken from
inclusive electron scattering \cite{DS99l}. This is so even in the
interval $\psi' < -1$ for all nuclei considered, in contrast to
the results of the RFG model where $f_{RFG}(\psi') = 0$ for $\psi'
\leq -1$. Here we emphasize that our scaling function $f(\psi')$
is obtained using the experimental information on the density
distribution. At the same time, however, $f(\psi')$ is related to
the NMD $n(p)$, as can be seen from Eq. (\ref{2.C.34}). We note
that Eqs.~(\ref{2.C.33}) and (\ref{2.C.34}) are equivalent when we
calculate in the CDFM the NMD $n(p)$ consistently using
Eq.~(\ref{2.C.11}), where the weight function $|F(R)|^2$ is
calculated using the derivative of the density distribution
$\rho(r)$ (Eq.~(\ref{2.C.16})). The same consistency exists in the
calculations of the CDFM Fermi momentum $k_F$ (which is used in
the calculations of $f(\psi')$ from Eqs. (\ref{2.C.33}) and
(\ref{2.C.34})). It is calculated by means of Eq.~(\ref{2.C.29})
and, equivalently, by Eq.~(\ref{2.C.31}) where the CDFM result for
$n(p)$ is used. The calculated values of $k_F$ in the CDFM are:
1.201 fm$^{-1}$ for $^4$He, 1.200 fm$^{-1}$ for $^{12}$C, 1.267
fm$^{-1}$ for $^{27}$Al, 1.270 fm$^{-1}$ and 1.200 fm$^{-1}$ for
$^{197}$Au.

\begin{figure}[thb]
\includegraphics[width=12cm]{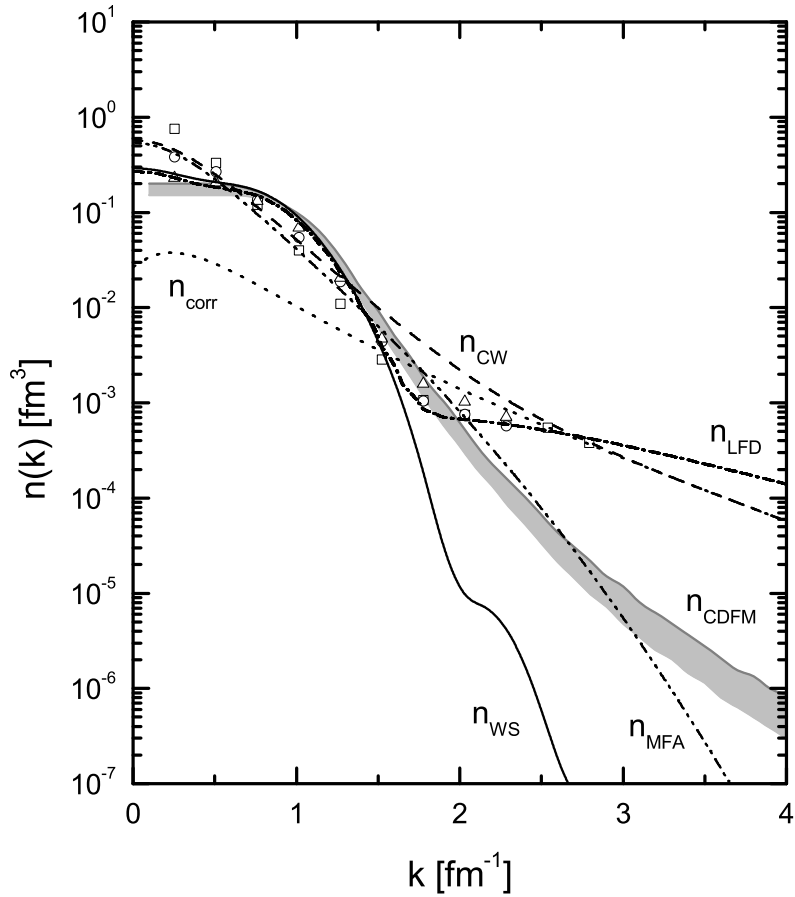}
\caption{Nucleon momentum distribution $n(k)$ from: i) CDFM: the
calculated results using Eqs.~(\protect\ref{2.C.11}) and
(\protect\ref{2.C.16}) for $^4$He, $^{12}$C, $^{27}$Al, $^{56}$Fe
and $^{197}$Au are combined by shaded area ($n_{CDFM}$); ii)
``$y$-scaling data'' \protect\cite{CPS91} given by open squares,
circles and triangles for $^4$He, $^{12}$C and $^{56}$Fe,
respectively; iii) $y_{CW}$-analyses \protect\cite{CW99,CW97} for
$^{56}$Fe [Eq.~(\protect\ref{2.A.9})] ($n_{CW}$), $n_{MFA}$
[Eq.~(\protect\ref{2.A.10})], $n_{corr}$
[Eq.~(\protect\ref{2.A.11})]; iv) LFD approach
\protect\cite{AGI+02}
[Eqs.~(\protect\ref{III.1})-(\protect\ref{III.4}),
(\protect\ref{III.6}), (\protect\ref{III.7})] for $^{56}$Fe
($n_{LFD}$); v) MFA calculations using Woods-Saxon single-particle
wave functions for $^{56}$Fe ($n_{WS}$).}
\label{fig2}
\end{figure}

In Fig.~\ref{fig2} we give the results of the calculations of the
NMD $n(k)$ within the CDFM for $^4$He, $^{12}$C, $^{27}$Al,
$^{56}$Fe and $^{197}$Au using Eqs. (\ref{2.C.11}), (\ref{2.C.16})
and Fermi-density distribution $\rho$ with parameter values
mentioned above (with b = 1.0 fm for $^{197}$Au). The
normalization is $\displaystyle \int n(\mathbf{k})d\mathbf{k}=1$.
The $n(k)$ from CDFM for the nuclei considered are with similar
tails at $k\gtrsim 1.5$ fm$^{-1}$, so they are combined and
presented by a shaded area. As can be expected, this similarity of
the high-momentum components of $n(k)$ leads to the superscaling
phenomenon. In our work there is an explicit relation of the
scaling function to the NMD in finite nuclear systems
[Eq.~(\ref{2.C.11})]. As can be seen, when the latter is
calculated in a realistic nuclear model accounting for nucleon
correlations beyond the MFA, a reasonable explanation of the
superscaling behavior of the scaling function for $\psi'< -1$ is
achieved.

In Fig.~\ref{fig2} we give also: i) the ``$y$-scaling data'' for
$n(k)$ in $^4$He, $^{12}$C and $^{56}$Fe obtained from analyses of
$(e,e')$ cross sections in \cite{CPS91} on the basis of the
$y$-scaling theoretical scheme, ii) $n(k)$ calculated within the
MFA using Woods-Saxon single-particle wave functions for
$^{56}$Fe; iii) the NMD for $^{56}$Fe [Eq.~(\ref{2.A.9})]
extracted from the more recent $y$-scaling analyses in
\cite{CW99,CW97}. We give also the corresponding contributions to
$n(k)$, namely the mean-field one $n_{MFA}(k)$
[Eq.~(\ref{2.A.10})] and the high-momentum component $n_{corr}(k)$
[Eq.~(\ref{2.A.11})], iv) the NMD $n(k)$, e.g. for $^{56}$Fe
obtained within an approach \cite{AGI+02} based on the NMD in the
deuteron from the light-front dynamics (LFD) method (e.g.
\cite{CK95,CDK98} and references therein). In \cite{AGI+02} $n(k)$
was written within the natural-orbital representation \cite{Low55}
as a sum of hole-state ($n^h(k)$) and particle-state ($n^p(k)$)
contributions
\begin{equation}\label{III.1}
n(k)=N_A\left[ n^h(k)+ n^p(k)\right] .
\end{equation}
In (\ref{III.1})
\begin{equation}\label{III.2}
n^h(k)= \sum_{nlj}^{F.L.} 2(2j+1) \lambda_{nlj} C(k)
|R_{nlj}(k)|^2 ,
\end{equation}
where $F.L.$ denotes the Fermi-level, and
\begin{equation}\label{III.3}
C(k)= \frac{m_N}{(2\pi)^3\sqrt{k^2+m_N^2}} .
\end{equation}
In Eq.~(\ref{III.2}) $\lambda_{nlj}$ are the natural occupation
numbers (which for the hole-states are close to unity and were set
to be equal to unity in \cite{AGI+02} with good approximation) and
the hole-state natural orbitals $R_{nlj}(k)$ are replaced by
single-particle wave functions from the MFA. In \cite{AGI+02}
Woods-Saxon single-particle wave functions were used for protons
and neutrons. The use of other s.p. wave functions (e.g. from
Hartree-Fock-Bogolyubov calculations) leads to similar results.

The normalization factor has the form:
\begin{equation}\label{III.4}
N_A= \left\{ 4\pi \int_0^\infty dq \, q^2 \left[ \sum_{nlj}^{F.L.}
2(2j+1) \lambda_{nlj} C(q) |R_{nlj}(q)|^2 + \frac{A}{2} n_5(q)
\right] \right\}^{-1} .
\end{equation}
The well-known facts: i) that the high-momentum components of
$n(k)$ caused by short-range and tensor correlations are almost
completely determined by the contributions of the particle-state
natural orbitals (e.g. \cite{SAD93}), and ii) the approximate
equality of the high-momentum tails of $n(k)/A$ for all nuclei
which are rescaled version of the NMD in the deuteron $n_d(k)$
\cite{FCW00}:
\begin{equation}\label{III.5}
n_A(k)\simeq \alpha_A n_d(k)
\end{equation}
(where $\alpha_A$ is a constant) made it possible to assume in
\cite{AGI+02} that $n^p(k)$ is related to the high-momentum
component $n_5(k)$ of the deuteron:
\begin{equation}\label{III.6}
n^p(k)= \frac{A}{2} n_5(k) .
\end{equation}
In (\ref{III.4}) and (\ref{III.6}) $n_5(k)$ is expressed by an
angle-averaged function \cite{AGI+02}
\begin{equation}\label{III.7}
n_5(k)= C(k) \overline{(1-z^2) f_5^2(k)} .
\end{equation}
In Eq. (\ref{III.7})
$z=\cos(\widehat{\mathbf{k}},\widehat{\mathbf{n}})$,
$\widehat{\mathbf{n}}$ being a unit vector along the 3-vector
($\overrightarrow{\omega}$) component of the four-vector $\omega$
which determines the position of the light-front surface
\cite{CK95,CDK98}. The function $f_5(k)$ is one of the six scalar
functions $f_{1-6}(k^2,\mathbf{n}\cdot\mathbf{k})$ which are the
components of the deuteron total wave function
$\Psi(\mathbf{k},\mathbf{n})$. The component $f_5$ exceeds
sufficiently other $f$-components for $k\geq 2 \div 2.5$ fm$^{-1}$
and is the main contribution to the high-momentum component of
$n_d(k)$, incorporating the main part of the short-range features
of the nucleon-nucleon interaction.

As can be seen in Fig.~\ref{fig2} the calculated LFD momentum
distributions are in a good agreement with the ``$y$-scaling
data'' for $^4$He, $^{12}$C and $^{56}$Fe from \cite{CPS91},
including the high-momentum region. We emphasize that $n(k)$
calculated in the LFD method does not contain any free parameters.

The comparison of the NMD's from the CDFM, LFD and from the
$y$-scaling analysis (YS) [Eqs.~(\ref{2.A.9})-(\ref{2.A.11})]
shows their similarity for momenta $k \lesssim 1.5$ fm$^{-1}$
(i.e. in the region where the MFA is a good approximation). It
also shows their quite different decreasing slopes for $k > 1.5$
fm$^{-1}$, where the effects of nucleon correlations dominate. In
the rest of this Section we will try to consider in more details
the questions concerning the reliability of the information about
the NMD obtained from the $y$- and $\psi'$-scaling analyses. This
concerns the sensitivity of such analyses and also the
identification of the intervals of momenta in which $n(k)$ can be
obtained with more reliability from the experimental data and from
the $y$- and $\psi'$-scaling studies. First of all, we emphasize
that the approaches considered to obtain experimental information
on $n(k)$ are strongly model dependent. In this respect we note
various ways to introduce the scaling variables, e.g. the
different $y$-scaling variables \cite{CPS91,CW99,CW97}, the
different $\psi$-scaling variables (e.g. in \cite{AMD+88,DS99}),
as well as the corresponding $y$- and $\psi$-scaling functions.
Even so, however, it is nevertheless worth considering in more
detail the model-dependent empirical information about the NMD
coming from $y$- and $\psi$-scaling analyses. We emphasize that
our consideration is based on $f(\psi')$ and the $y$-scaling
function $F(y)$ within the CDFM, as well as on the relationship
between both of them discussed in Subsection~\ref{ss:d}.

\begin{figure}[th]
\includegraphics[width=12cm]{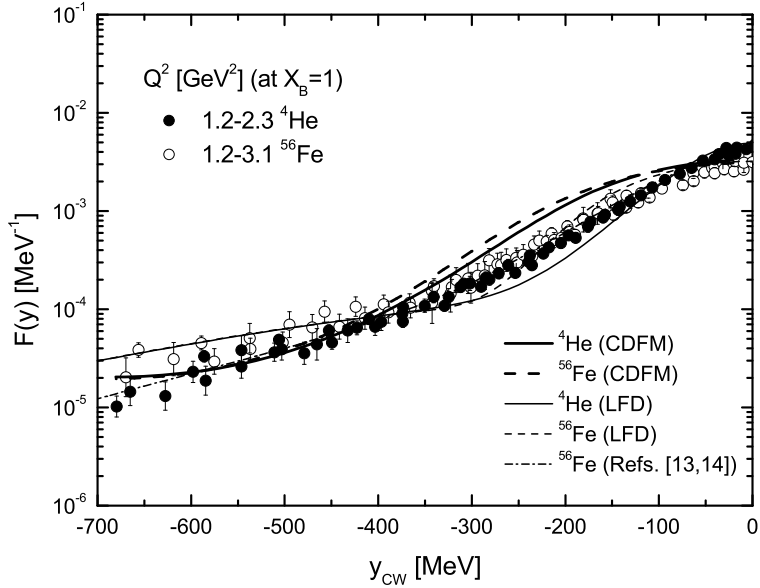}
\caption{The $y$-scaling function $F(y)$ for $^4$He and $^{56}$Fe
calculated in the CDFM [Eq.~(\ref{2.D.3})] (solid and thick dashed
lines), from the $y_{CW}$-scaling approach
\protect\cite{CW99,CW97} [Eqs.~(\ref{2.A.9})-(\ref{2.A.11})] for
$^{56}$Fe (dash-dotted line) and from the approach
\protect\cite{AGI+02} within the LFD method
[Eqs.~(\protect\ref{III.1})-(\protect\ref{III.4}),
(\protect\ref{III.6}), (\protect\ref{III.7})] (thin solid and
dashed lines). The results are compared with the $y_{CW}$-scaling
data taken from \protect\cite{CW99,CW97}.}
\label{fig3}
\end{figure}

Firstly, we give the results of the calculations of the
$y$-scaling function $F(y)$ obtained in the CDFM
[Eq.~(\ref{2.D.3})] using different NMD $n(k)$: i) from the CDFM
(Eqs.~(\ref{2.C.11}) and (\ref{2.C.16})), ii) from the $y_{CW}$
scaling approach \cite{CW99,CW97}
[Eqs.~(\ref{2.A.9})-(\ref{2.A.11})], and iii) from the approach
\cite{AGI+02} which uses the results of the LFD method
[Eqs.~(\ref{III.1})-(\ref{III.4}), (\ref{III.6}), (\ref{III.7})].
In Fig.~\ref{fig3} they are compared with the $y_{CW}$ scaling
data for $F(y)$ for the $^4$He and $^{56}$Fe nuclei taken from
\cite{CW99,CW97}. As can be seen from Fig.~\ref{fig3}, there is a
general agreement with the data for all the NMD's considered. At
first thought this can be surprising knowing the different
behavior of $n(k)$'s for larger $k$ that are seen in
Fig.~\ref{fig2}. The reasons for the relative similarity of the
results for $F(y)$ using different $n(k)$'s are as follows.

i) The $\psi'$- scaling variable is a quadratic function of $y$
but not a linear one [Eq.~(\ref{2.B.7})]. In accordance with this,
the lower limit of the integral [Eq.~(\ref{2.D.3})] for $F(y)$ is
not $|y|$ as in \cite{CW99,CW97}, but $|y|(1-c|y|)$ (see for a
comparison Eqs. (\ref{2.D.6}) and (\ref{2.D.7})); ii) Due to the
steep slope rates of decreasing of the NMD's for large momenta,
the main contribution to the integral (\ref{2.D.3}) (and to the
estimation (\ref{2.D.6})) comes from momenta which are not much
larger than the lower limit of the integration. In this way, the
very high-momentum components of $n(k)$ do not play so important a
role (in the integral in (\ref{2.D.3})), at least for momenta
studied so far $y>-700$ MeV/c. We give some numerical estimations:
for example, for $y=-300$ MeV/c instead of integrating from $|y|=
300$ MeV/c $=1.52$ fm$^{-1}$ (as in (\ref{2.D.7})), in $F(y)$ in
(\ref{2.D.6}) the integration starts from $|y|(1- c|y|)=1.19$
fm$^{-1}$, for $y=-600$ MeV/c instead of integrating from $|y|=
600$ MeV/c $= 3.04$ fm$^{-1}$ the lower limit of the integral in
(\ref{2.D.6}) is $|y|(1- c|y|)=1.71$ fm$^{-1}$. This means that
the main contribution to $F(y)$ from $n(k)$ is for momenta
$k\lesssim 2$ fm$^{-1}$. The behavior of $F(y)$ in Fig.~\ref{fig3}
reflects that one of $n(k)$. For instance, for $-400\leq y\leq 0$
MeV/c the CDFM result for $F(y)$ is higher than those of LFD and
YS because the values of $n(k)$ from CDFM  for $k\leq 1.5$
fm$^{-1}$ are larger than those of $n(k)$ from the LFD and the YS.
In contrast to this, the values of $F(y)$ for $-700\leq y\leq
-400$ MeV/c in the CDFM result are lower than those of the LFD
because $n(k)$ from LFD has a higher tail than $n(k)$ in the CDFM
for $k>1.5$ fm$^{-1}$. Nevertheless, though the tails of $n(k)$
for large $k$ are quite different (for $k>1.5$ fm$^{-1}$), the
values of $F(y)$ from the different approaches are quite close to
each other and are in agreement with the existing data. In this
way, we can conclude from our experience that the existing
$y$-scaling data can give reliable information for the NMD for
momenta not larger than $1.5 \div 2.0$ fm$^{-1}$, where the
considered $n(k)$ are not drastically different from each other.

One can see from Fig.~\ref{fig3} that the CDFM results for $F(y)$ are
in a very good agreement with the data for $^4$He for $y\lesssim
-400$ MeV/c, while in the same region the result of the LFD agrees
very well with the data for $^{56}$Fe. The YS result for $F(y)$
agrees well with the data for $^{56}$Fe for $y\gtrsim -600$ MeV/c.

It is worth mentioning that in our approach we start from the
$\psi'$-scaling consideration for the function $F(y)$ and this
leads to a relatively good description of the $y$-scaling data on
the basis of the correct accounting of the relationship between
the $\psi'$- and $y$-scaling variables. The overall agreement of
the theoretical results using the momentum distributions from the
CDFM, the LFD and the YS with the experimental data for $F(y)$ is
related with their similarities up to momenta $k= 1.5 \div 2.0$
fm$^{-1}$.

\begin{figure}[th]
\includegraphics[width=10cm]{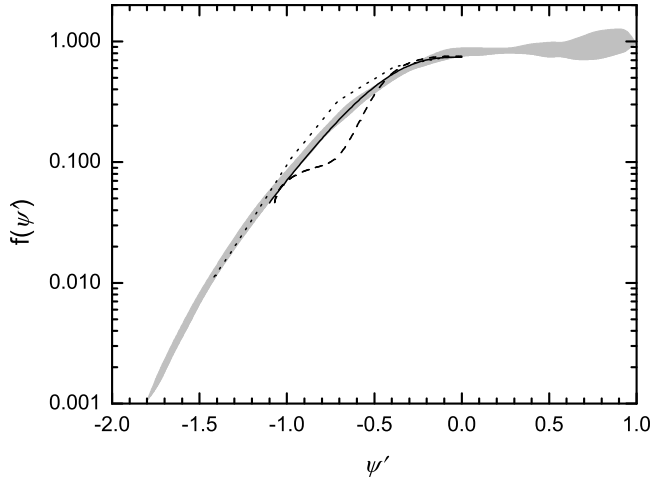}
\caption{The $\psi'$-scaling function $f(\psi')$ at $^{56}$Fe and
$q=1000$ MeV/c calculated from Eq.~(\ref{2.D.13}) using $n(k)$
from: i) the $y_{CW}$-scaling analysis \protect\cite{CW99,CW97}
[Eqs.~(\ref{2.A.9})-(\ref{2.A.11})] (solid line); ii) the approach
\protect\cite{AGI+02} within the LFD method
[Eqs.~(\protect\ref{III.1})-(\protect\ref{III.4}),
(\protect\ref{III.6}), (\protect\ref{III.7})] (dashed line). The
CDFM result obtained using Eqs.~(\protect\ref{2.C.25}) and
(\protect\ref{2.C.26}) is given by dotted line. The experimental
data given by shaded area are taken from \protect\cite{DS99}.}
\label{fig4}
\end{figure}

Our next step is to estimate the $\psi'$-scaling function
$f(\psi')$ [Eqs.~(\ref{2.C.40}) and (\ref{2.D.8})] replacing the
lower limit of the integration $k_F|\psi'|$ approximately by $|y|$
which is, however, a solution of (\ref{2.D.12}), i.e.
$|y|=\displaystyle \frac{1}{2c}\left( 1- \sqrt{1-4ck_F
|\psi'|}\right)$, but is not the linear function of $|\psi'|$:
$|y| = k_F |\psi '|$. This is done in order to introduce in the
relationship of $f(\psi')$ with $F(y)$ in (\ref{2.D.8}) and
(\ref{2.D.10}) the lower limit in the integral for $F(y)$ to be
$|y|$ (as in the YS) where, however, the correct relationship of
$|y|$ with $|\psi'|$ (Eq.~(\ref{2.D.12})) is accounted for. In
Fig.~\ref{fig4} we give the results for $f(\psi')$ from
Eq.~(\ref{2.D.13}) using the NMD from the YS analysis
\cite{CW99,CW97} [Eqs.~(\ref{2.A.9})-(\ref{2.A.11})] and from the
approach \cite{AGI+02} within the LFD method
[Eqs.~(\ref{III.1})-(\ref{III.4}), (\ref{III.6}), (\ref{III.7})].
One can see that the NMD from the YS analysis
[Eqs.~(\ref{2.A.9})-(\ref{2.A.11})] gives a good description of
$f(\psi')$ for $^{56}$Fe in the case of $q=1000$ MeV/c for values
of $\psi'$: $-1.10 \leq \psi' \leq 0$ (for which $y<0$ and $|y|
\leq 1/(2c)$ at $c=0.144$ fm). The scaling function $f(\psi')$
calculated by $n(k)$ from the LFD is in agreement with the data
for $-0.5 \lesssim \psi' \leq 0$, while in the region $-1.1 \leq
\psi' \leq -0.5$ shows a dip in the interval $-0.9< \psi' \leq
-0.6$. The difference in the behavior of $f(\psi')$ in these two
cases reflects the difference of the momentum distributions of YS
and LFD in the interval $1.5 \lesssim k \lesssim 2.5$ fm$^{-1}$:
the $n(k)$ of the LFD has a dip around $k \approx 1.7$ fm$^{-1}$
below the curve of $n(k)$ from the YS analysis.

\section{conclusions \label{s:con}}

The results of the present work can be summarized as follows.

(i) The main aim of our work was to study the nucleon momentum
distributions from the experimental data on inclusive electron
scattering from nuclei which have shown the phenomenon of
superscaling. For this purpose we made an additional extension of
the coherent density fluctuation model in order to express the
$\psi'$-scaling function, $f(\psi')$, explicitly in terms of the
nucleon momentum distribution for realistic finite systems. This
development is a natural extension of the relativistic Fermi gas
model. In this way $f(\psi')$ can be expressed equivalently by
means of both density and momentum distributions. In \cite{AGK+04}
our results on $f(\psi')$ were obtained on the basis of the
experimental data on the charge densities for a wide range of
nuclei. In the present work we discuss the properties of $n(k)$
which correspond to the results for $f(\psi')$ obtained in the
CDFM. Thus we show how both quantities, the density and the
momentum distribution, are responsible for the scaling behavior in
various nuclei.

(ii) In addition to the work presented in Ref. \cite{AGK+04}, the
scaling function $f(\psi')$ is calculated here in the CDFM at $q=
1560$ MeV/c. The comparison with the data from \cite{DS99l} shows
superscaling for negative values of $\psi'$ including $\psi' < -
1$, in contrast to the RFG model where $f(\psi') = 0$ for $\psi'
\leq -1$.

(iii) The $y$-scaling function $F(y)$ is defined in the CDFM on
the basis of the RFG relationships. The calculations of $F(y)$ are
performed in the model using three different momentum
distributions: from the CDFM, from the $y$-scaling analyses
\cite{CW99,CW97} and from the theoretical approach based on the
light-front dynamics method \cite{AGI+02}. Comparing the results
of the calculations for $^4$He and $^{56}$Fe nuclei with the
experimental data, we show the sensitivity of the calculated
$F(y)$ to the peculiarities of the three $n(k)$'s in different
regions of momenta.

(iv) An approximate relationship between $f(\psi')$ and $F(y)$ is
established. It is shown that the momentum distribution $n_{CW}$
for $^{56}$Fe from the $y$-scaling studies in \cite{CW99,CW97} can
describe to a large extent the empirical data on $f(\psi')$ for
$q=1000$ MeV/c. We point out that the interrelation and the
comparison between the results of the $\psi'$- and $y$-scaling
analyses have to be studied accounting for the correct non-linear
dependence of $\psi'$ on the $y$-scaling variable, which reflects
the dependence on the missing energy.

(v) The regions of momenta in $n(k)$ which are mainly responsible
for the description of the $y$- and $\psi'$-scaling are estimated.
It is shown in the present work that the existing data on the $y$-
and $\psi'$-scaling are informative for the momentum distribution
$n(k)$ at momenta up to $k \lesssim 2 \div 2.5$ fm$^{-1}$. It can
be concluded that further experiments are necessary for studies of
the high-momentum components of the nucleon momentum distribution.

\acknowledgments One of the authors (A.N.A.) is grateful for warm
hospitality to the Faculty of Physics of the Complutense
University of Madrid, to the Instituto de Estructura de la
Materia, CSIC., and for support during his stay there to the State
Secretariat of Education and Universities of Spain
(Nº.Ref.SAB2001-0030). Four of the authors (A.N.A., M.K.G., D.N.K.
and M.V.I.) are thankful to the Bulgarian National Science
Foundation for partial support under the Contracts Nos. $\Phi$-905
and $\Phi$-1416. This work was partly supported by funds provided
by DGI of MCyT (Spain) under Contracts BFM 2002-03562, BFM
2000-0600 and BFM 2003-04147-C02-01 and by the Agreement (2004
BG2004) between the CSIC (Spain) and the Bulgarian Academy of
Sciences.


\end{document}